\documentclass[12pt,aps,prb]{revtex4}
\usepackage{epsfig}
\usepackage{amsmath}

\normalsize

\def\bj{{\bf j}}
\def\bk{{\bf k}}

\def\bp{{\bf p}}
\def\bq{{\bf q}}

\def\bQ{{\bf Q}}

\def\b0{{\bf 0}}

\def\Re{{\rm Re}}
\def\Im{{\rm Im}}

\def\up{\uparrow}
\def\down{\downarrow}

\def\eps{\epsilon}

\def\Gam{\Gamma}

\def\Lam{\Lambda}
\def\om{\omega}

\def\sg{\sigma}
\def\Sg{\Sigma}

\begin{document}


\title{Pseudogap at hot spots in the two-dimensional
 Hubbard model at weak coupling}

\author{Daniel Rohe and Walter Metzner}

\affiliation{Max Planck Institute for Solid State Research,
 D-70569 Stuttgart, Germany}

\date{\small\today}


\begin{abstract}
We analyze the interaction-induced renormalization of
single-particle excitations in the two-dimensional Hubbard model
at weak coupling using the Wick-ordered version of the
functional renormalization group.
The self energy is computed for real frequencies by
integrating a flow equation with renormalized two-particle
interactions.
In the vicinity of hot spots, that is points where the Fermi surface
intersects the umklapp surface, self energy effects beyond the usual 
quasi-particle renormalizations and damping occur near instabilities 
of the normal, metallic phase.
Strongly enhanced renormalized interactions between particles at 
different hot spots generate a pronounced low-energy peak in the 
imaginary part of the self energy, leading to a pseudogap-like 
double-peak structure in the spectral function for single-particle
excitations.

\noindent
\mbox{PACS: 71.10.Fd , 71.27.+a , 71.10.Hf}\\
\end{abstract}

\maketitle


\section{Introduction}

The most popular prototype model for the dynamics of electrons in
the copper oxide planes of cuprate superconductors is the
two-dimensional Hubbard model,\cite{And}
which captures two of the most prominent and universal
features of these compounds:
to be an antiferromagnetic Mott insulator at half filling and a
d-wave superconductor at sufficiently strong doping.
It is very difficult to compute the low-energy properties of
the two-dimensional Hubbard model at strong coupling, 
as relevant to the cuprates.
In fact, close to half filling the model exhibits complicated 
non-trivial behavior
already at weak coupling, due to a mutual interplay of
singularities in the particle-particle and particle-hole channel,
which cannot be captured by simple resummations of Feynman
diagrams.

A systematic analysis of this interplay has been initiated in 
recent years with the development of functional renormalization
group (fRG) methods for interacting Fermi systems.\cite{Sal}
The starting point of this approach is an exact hierarchy of 
differential flow equations for Green or vertex functions,
which is obtained by differentiating the generating functional 
of the system with respect to an infrared cutoff $\Lam$ or 
some other energy scale.
Approximations are then constructed by truncating this hierarchy
and parametrizing the vertex functions with a managable set of
variables or functions.
Already from the first fRG studies of the two-dimensional Hubbard 
model, which focussed on the flow of the two-particle vertex 
$\Gam$ and various susceptibilities in the one-loop approximation, 
a number of interesting conclusions could be drawn.\cite{ZS,HM,HSFR}
In particular, the expected existence of superconductivity with 
$d_{x^2-y^2}$ symmetry was firmly established, at least at 
weak coupling. More importantly, it became clear that close to 
half filling correlations in the particle-particle and particle-hole 
channel are both strong and mutually influence each other.

The effect of correlations on single-particle excitations is seen 
most clearly in the  self energy $\Sg(\bk,\om )$, for frequencies 
$\om$ on the real axis.
Zanchi \cite{Zan} computed the flow of the frequency derivative
of the self energy on the Fermi surface,
which determines the wave function renormalization factor 
$Z = [1 - \left.\partial_{\om}\Re\Sg(\bk_F,\om)\right|_{\om = 0}]^{-1}$,
for the special case of a half-filled Hubbard model with pure
nearest neighbor hopping. 
In Fermi liquids $Z$ yields the weight of the quasi-particle peak 
in the spectral function.
Zanchi found that the $Z$-factor is strongly reduced when the 
renormalized vertex increases, the suppression being strongest 
near the van Hove points.
Subsequently, Honerkamp and Salmhofer \cite{HS} computed the flow
of the $Z$-factor for the two-dimensional Hubbard model at as well 
as away from half filling.
While confirming an anisotropic reduction of $Z$, they observed
that this suppression emerges much slower than the divergence 
of dominant interactions. 
In addition, they pointed out that the self energy does not
significantly alter the vertex renormalization in the
perturbative regime, that is before the interactions grow strongly.
An fRG calculation of the imaginary part of the self energy at
zero frequency, $\Im\Sg(\bk_F,0)$, was performed by
Honerkamp.\cite{Hon}
In Fermi liquids this quantity is directly related to the decay rate
of quasi-particles.
For a hole-doped Hubbard model with a concave Fermi surface that 
almost touches the van Hove points he found that $\Im\Sg(\bk_F,0)$ 
is moderately anisotropic, with larger values for momenta closer 
the van Hove points. 
A closer look reveals that the imaginary part of the self energy 
is actually largest near a hot spot, foreshadowing already a main 
feature of our results presented below.
The first fRG calculation of the full frequency dependence of
the self energy has been performed very recently by Katanin
and Kampf.\cite{KK}
They computed the real and imaginary parts of $\Sg(\bk_F,\om)$
and the resulting spectral function $A(\bk_F,\om)$ for the
Hubbard model with nearest and next-to-nearest neighbor hopping
at temperatures above $T^*$
and observed the destruction of Fermi liquid behavior near $T^*$
due to structures indicating the formation of a pseudogap,
in particular for $\bk_F$ close to a van Hove point.
For $\bk_F$ at short but finite distance from a van Hove point
they found an additional peak between the two maxima that
confine the pseudogap in the spectral function. Moving even further 
away from the van Hove point toward the diagonal of the Brillouin
zone the three peaks gradually merge into one single peak.\cite{KK}

In this work we present a calculation of the full frequency
dependence of the self energy and the resulting spectral function
for the two-dimensional Hubbard model as obtained from the
\emph{Wick-ordered} version of the fRG.\cite{Sal,HM} This scheme
has the advantage that only low-energy states contribute to the
flow at low-energy scales, which leads to a better control of
the necessary Fermi surface projections of momentum dependences
compared to other fRG versions.\ \cite{HS,Hon,KK} 
We do the calculation for frequencies directly on the real axis, 
thus avoiding analytical continuation.
In agreement with previous studies we obtain marked deviations
from Fermi liquid behavior for Fermi momenta near
van Hove points, and also near other hot spots in the case of
filling factors above the van Hove singularity: the imaginary
part of the self energy develops a pronounced peak
at low frequencies, which leads to a double-peak structure in
the spectral function reminiscent of a pseudogap.
This double peak gradually transforms into a single peak when
moving away from the hot spot or van Hove regions, and also upon
raising the temperature.
Unlike Ref.\ \onlinecite{KK} we never obtain
a third peak in the pseudogap valley.

\section{Model}

We consider the one-band Hubbard model defined by the Hamiltonian
\begin{equation}
 H = \sum_{\bj,\bj'} \sum_{\sg}
 t_{\bj\bj'} \, c^{\dag}_{\bj\sg} c^{\phantom{\dag}}_{\bj'\sg} +
 U \sum_{\bj} n_{\bj\up} n_{\bj\down} ,
\end{equation}
with a local repulsion $U>0$ and hopping amplitudes
$t_{\bj\bj'} = -t$ between nearest neighbors and
$t_{\bj\bj'} = -t'$ between next-to-nearest neighbors on a
square lattice.
The corresponding dispersion relation
\begin{equation}
 \eps_{\bk} = -2t (\cos k_x + \cos k_y)
 - 4t' \cos k_x \cos k_y
\end{equation}
has saddle points at $\bk = (0,\pi)$ and $(\pi,0)$, leading
to logarithmic van Hove singularities in the non-interacting
density of states at energy $\eps_{\rm vH} = 4t'$.

\section{Method}

We use the Wick-ordered version of the fRG as proposed by Salmhofer,\ 
\cite{Sal} which has been concisely described and used in the 
context of an explicit one-loop
computation of the two-particle vertex for the two-dimensional
Hubbard model.\ \cite{HM}
In this scheme, to second order in the two-particle vertex the flow 
of the self energy $\Sg^{\Lam}$ is determined by a single two-loop
Feynman diagram, shown in Fig.\ 1, and the flow of the
two-particle vertex $\Gam^{\Lam}$ is obtained to second
order by evaluating the one-loop diagram in Fig.\ 1.
The internal lines without slash in the Feynman diagrams
correspond to the bare propagator
\begin{equation}
 D^{\Lam}(k) = \frac{\Theta(\Lam - |\xi_{\bk}|)}
 {ik_0 - \xi_{\bk}} \; ,
\end{equation}
where $\xi_{\bk} = \eps_{\bk} - \mu$ and $\Lam > 0$ is the
cutoff; the slashed lines represent
$\partial_{\Lam} D^{\Lam}$, which is proportional to
$\delta(\Lam - |\xi_{\bk}|)$.
Note that $D^{\Lam}$ has support only on a shell of energetic
width $\Lam$ around the Fermi surface, that is for momenta
$\bk$ with $|\xi_{\bk}|$ \emph{below} scale $\Lam$.
A first order contribution to $\Sg^{\Lam}$ (Hartree term)
has been absorbed in a shift of the chemical potential, such
that the leading contributions are quadratic in $\Gam^{\Lam}$.
We note that in the full hierarchy of flow equations of the
Wick-ordered fRG also one-particle reducible terms contribute.
However, these have to be considered only if the flow is
computed beyond second order in $\Gam^{\Lam}$.

For a spin-rotationally invariant system the
vertex is uniquely determined by two functions depending on momenta
only, namely a singlet component $\Gam_s^{\Lam}$ and a triplet 
component $\Gam_t^{\Lam}$, as described in Ref.\ \onlinecite{HM}. 
In the present work the vertex is defined with an extra factor
$1/2$ compared to the conventions in Ref.\ \onlinecite{HM}.
The flow equation for the vertex can only be solved if the momentum
and frequency dependence of the vertex is parametrized by a
reduced set of variables.
The frequency dependence of $\Gam^{\Lam}$ can be neglected without
much damage because it is absent in the bare interaction, and
irrelevant in the sense of power counting in the limit $\Lam \to 0$.
The frequency integrals in the Feynman diagrams in Fig.\ 1 can
then be carried out analytically.
The resulting one-loop flow equations for
$\Gam_s^{\Lam}(\bk'_1,\bk'_2;\bk_1,\bk_2)$ and
$\Gam_t^{\Lam}(\bk'_1,\bk'_2;\bk_1,\bk_2)$ given in Ref.\
\onlinecite{HM} are supplemented by the flow equation for the 
imaginary part of $\Sg^{\Lam}(\bk,\om)$ for
arbitrary real frequencies $\om$ given as
\begin{eqnarray}
 \partial_{\Lam} \Im\Sg^{\Lam}(\bk,\om) &=&
 \pi \int \frac{d^2{\mathbf p}}{(2\pi)^2} \int \frac{d^2{\mathbf q}}{(2\pi)^2} \,
 \partial_{\Lam} \big[
 \Theta(\Lam - |\xi_{\bp}|) \,
 \Theta(\Lam - |\xi_{\bq}|) \,
 \Theta(\Lam - |\xi_{\bk+\bq-\bp}|) \big] \nonumber\\[2mm]
 &\times& \big\{
   [\Gam_s^{\Lam}(\bk,\bq;\bp,\bk+\bq-\bp)]^2 +
  3[\Gam_t^{\Lam}(\bk,\bq;\bp,\bk+\bq-\bp)]^2 \big\}
 \nonumber\\[2mm]
 &\times& F(\bq;\bp,\bk\!+\!\bq\!-\!\bp) \,
 \delta(\om - \xi_{\bp} + \xi_{\bq} - \xi_{\bk+\bq-\bp})
\end{eqnarray}
with the combination of Bose ($b$) and Fermi ($f$) functions
\begin{equation}
 F(\bq,\bp,\bp') =
 [ f(\xi_{\bq}) - f(\xi_{\bp'}) ] \,
 [ f(\xi_{\bp}) + b(\xi_{\bq}-\xi_{\bp'}) ] \; .
\end{equation}
In contrast to its appearance $F$ is symmetric in $\bp$ and $\bp'$.
The real part of $\Sg^{\Lam}$ can be determined from $\Im\Sg^{\Lam}$
via a Kramers-Kronig relation.
The momentum dependence of the vertex is discretized similarly to previous
fRG calculations \cite{ZS,HM,HSFR} by dividing the Brillouin zone
into patches as shown in Fig.\ 2.
The shape of the patches accounts for the fact that the
momentum dependence normal to the Fermi surface is irrelevant in
the sense of power counting, in contrast to the tangential momentum
dependence.
The flow equations are integrated from
$\Lam = \Lam_0 = \max_{\bk} |\xi_{\bk}|$ to $\Lam = 0$.
The initial conditions are $\Gam_s^{\Lam_0} = U$,
$\Gam_t^{\Lam_0} = 0$, and $\Sg^{\Lam_0} = 0$.
From the result for the self energy $\Sg^{\Lam}$ for $\Lam \to 0$
we compute the spectral function $A(\bk,\om) = -2\Im G(\bk,\om)$.
To avoid the computationally expensive self-consistent determination
of the interacting Fermi surface, we remove Fermi surface shifts
by subtracting $\Re\Sg(\bk_F,0)$ from the self energy in the Dyson
equation $G^{-1} = G_0^{-1} - \Sg \,$.

\section{Results}

We focus on the two-dimensional Hubbard model
with a negative next-to-nearest neighbor hopping amplitude
$t' = -0.1t$ and two different average densities, $n=0.92$
and $n=1$, respectively.
In both cases the Fermi surface given by $\xi_{\bk}=0$ is
closed around $(\pi,\pi)$ and has finite curvature everywhere,
see Fig.\ 2.
In the following the so-called hot spots, that is the
points where the Fermi surface intersects the umklapp surface,
will play a special role.
The momenta on the umklapp surface obey the relation
$\eps_{\bk+\bQ} = \eps_{\bk}$ with $\bQ = (\pi,\pi)$.
For $n=0.92$ the Fermi surface almost touches the van Hove
points, and the hot spots are very close to the
latter. By contrast, for $n=1$ the hot spots are well separated
from the van Hove points.
We set $t=1$, that is all energies are given
in units of $t$.
We fix a low finite temperature $T = 0.05$ and choose the bare
interaction $U$ such that the couplings flow to strong values
but do not diverge for $\Lam \to 0$. In other words, the
temperature $T$ is slightly above the scale $T^*$ at which the
one-loop flow diverges.

The complete frequency dependence of the imaginary part of the
self energy $\Im\Sg(\bk_F,\om)$ for high and low energies is
shown in Fig.\ 3, for $n = 0.92$ and a Fermi momentum $\bk_F$
which coincides with a hot spot.
The fRG result is compared to the result from standard second
order perturbation theory, which is obtained from
the flow equation (4) when the renormalized vertex $\Gam^{\Lam}$
is replaced by the bare interaction. While the vertex
renormalization has virtually no influence on the self energy
at high frequencies, it leads to a pronounced peak in $\Im\Sg$ at 
low frequencies, which is absent in perturbation theory.

We now focus on the low-energy behavior. In Fig.\ 4 we plot
$\Sg(\bk_F,\om)$ at small frequencies for $n=0.92$ and six different
Fermi momenta $\bk_F$ situated in patches $1,2,\dots,6$,
respectively. We show two different fRG results: in one case
the bare interaction $U$ has been tuned such that the
renormalized vertex reaches a peak value
$\Gam^{\Lam=0}_{\rm max} = 15$, in the other
$\Gam^{\Lam=0}_{\rm max} = 150$.
If the peak in $\Im\Sg(\bk_F,\om)$ observed already in Fig.\ 3 
is sufficiently strong, it causes a sign change in the slope of $\Re\Sg(\bk_F,\om)$ at small 
$\om$. For $\left.\partial_{\om}\Re\Sg(\bk_F,\om)\right|_{\om = 0} > 1$ this leads to a double-peak structure reminiscent of a 
pseudogap in the spectral function $A(\bk_F,\om)$, shown in Fig.\ 5.
For a less strongly renormalized vertex, as well as for momenta 
away from the hot spot, the peak in $\Im\Sg(\bk_F,\om)$ disappears
gradually. 
As a consequence, the double peak in the spectral function is well
developed only for $\bk_F$ near the hot spot, and for a comparatively
large renormalized vertex, that is for $T$ rather close to $T^*$.
Otherwise, the self energy merely leads to an angle-dependent 
lifetime broadening of the quasi-particle peak in $A(\bk_F,\om)$.

For electron density $n=1$ the hot spot is situated in the
third patch (see Fig.\ 2). Results for $\Sg(\bk_F,\om)$
and $A(\bk_F,\om)$ are shown in Figs.\ 6 and 7, respectively.
The fRG again yields a peak in $\Im\Sg(\bk_F,\om)$ around
$\om = 0$ for a hot spot momentum $\bk_F$, but only
for $\Gam^{\Lam}_{\rm max} = 150$. Moreover, its absolute value is 
much smaller compared to the case $n=0.92$, and it is less 
pronounced. Consequently, the corresponding spectral function 
exhibits only a mild dip. For
weaker interactions or $\bk_F$ away from the hot spot
$A(\bk_F,\om)$ has a more or less conventional shape, with a single
lifetime-broadened peak. This broadening is larger for momenta next to the hot spot, since the low-frequency part of
$\Im\Sg(\bk_F,\om)$ is still strongly enhanced, albeit not peaked, in these regions.

Note that there is nothing special
about the density $n=1$ (half-filling) at weak coupling if
$t' \neq 0$, in contrast to the case of strong coupling,
for which the half-filled Hubbard model becomes a Mott insulator,
and in contrast to the special case with pure nearest neighbor
hopping ($t'=0$), for which the Fermi surface coincides with the
umklapp surface at half-filling. The fact that the anomaly we 
observe in the self energy is more pronounced for a Fermi surface 
close to van Hove filling reflects the relevance of the van Hove 
singularities in the weak coupling limit.

To understand the mechanism behind the low-energy singularity
in $\Im\Sg(\bk,\om)$ for hot spot momenta $\bk_{\rm hs}$,
we consider the flow eq.\ (4) in the limit
$T \to 0$ and $\om \to 0$.
At $T=0$ the function $F(\bq,\bp,\bp')$ reduces to a step
function which constrains the relative signs of $\xi_{\bq}$,
$\xi_{\bp}$, and $\xi_{\bp'}$:
\begin{equation}
 F(\bq,\bp,\bp') = \left\{ \begin{array}{rcl}
 -1 & {\rm for} & \xi_{\bq} < 0 \, ; \; \xi_{\bp},\xi_{\bp'} > 0 \\
 -1 & {\rm for} & \xi_{\bq} > 0 \, ; \; \xi_{\bp},\xi_{\bp'} < 0 \\
  0 & {\rm else}
 \end{array} \right.
\end{equation}
Energy conservation imposed by the $\delta$-function in Eq.\ (4)
then implies that in the limit $\om \to 0$ the internal momenta
$\bq$, $\bp$, $\bp'=\bk+\bq-\bp$ are all constrained to the
Fermi surface. If in addition $\bk$ is a Fermi momentum, all 
ingoing and outgoing momenta of $\Gam^{\Lam}$ in Eq.\ (4)
have to lie on the Fermi surface.
At low finite $T$ or $\om$ this constraint is slightly relaxed such
that the internal momenta can be taken from a thin shell of
energetic width of order $T$ or $\om$ around the Fermi surface.
Since one of the internal momenta $\bp$ on the right hand side of the
flow equation is forced to sit at scale $|\xi_{\bp}| = \Lam$ by the 
$\Lam$-derivative
of the cutoff functions, the flow of $\Im\Sg^{\Lam}$ receives
sizable contributions only when $\Lam$ has reached the scale $T$
or $\om$.
In two dimensions there are two generic types of momentum-conserving
interaction processes with all ingoing ($\bk_1,\bk_2$)
and outgoing ($\bk'_1,\bk'_2$) momenta on the Fermi surface:
forward scattering processes
($\bk'_1 = \bk_1$ or $\bk'_1 = \bk_2$) and Cooper processes ($\bk_2 = -\bk_1$). In lattice systems
also certain umklapp processes in which a reciprocal
lattice vector is added to the total momentum can fulfill this condition. They
can play an important role, as discussed below.  

Looking at the renormalized interactions obtained
from the one-loop flow of $\Gam^{\Lam}$ for our two choices of
densities, $n=0.92$ and $n=1$, we find that
in both cases the dominant effective interactions are in
the spin singlet channel with ingoing and outgoing momenta near
hot spots. Among these, interactions with a momentum transfer
$\bQ = (\pi,\pi)$ are leading.
For $n=0.92$ the vertex is maximal for the scattering of a
Cooper pair with both partners on hot spots to two other hot
spots with a momentum transfer $\bQ$, that is
$(\bk_{\rm hs},-\bk_{\rm hs}) \mapsto
 (\bk'_{\rm hs},-\bk'_{\rm hs})$
with $\bk'_{\rm hs} - \bk_{\rm hs} = \bQ$.
Combined Cooper and forward scattering processes
$(\bk_{\rm hs},-\bk_{\rm hs}) \mapsto
 (\bk_{\rm hs},-\bk_{\rm hs})$ are also among the strongly
enhanced interactions at that density.
Computing the contributions to $\Im\Sg$ from different 
scattering processes separately reveals that the peak at hot
spots is indeed generated mostly by Cooper and $\bQ$-transferring 
interactions, where the latter are slightly more important.
For $n=1$ the leading interaction is an umklapp process,
$(\bk_{\rm hs},\bk_{\rm hs}) \mapsto
 (\bk'_{\rm hs},\bk'_{\rm hs})$
with $\bk'_{\rm hs} - \bk_{\rm hs} = \bQ$,
closely followed by a $\bQ$-transferring Cooper process on hot 
spots. However, the total contribution from Cooper processes
(including the $\bQ$-transferring one) to the peak in $\Im\Sg$ 
is very small compared to the total contribution from 
$\bQ$-transferring processes in that case.
The importance of $\bQ$-transferring and Cooper interactions
in the two-dimensional Hubbard model is well known
from earlier fRG and other studies, and reflects strong
antiferromagnetic and superconducting correlations, respectively.

The above mechanism for the destruction of fermionic quasi-particles 
close to the Fermi surface is completely different from the
mechanism leading to Luttinger liquid behavior in one-dimensional
electron systems.\cite{Gia}
In one-dimensional (gapless) electron liquids the renormalized
interactions remain finite, while the self-energy receives 
nevertheless enhanced contributions at low energy due to a
large phase space for scattering processes.
However, $\Im\Sg(k_F,\om)$ does not develop a peak at $\om=0$,
it just vanishes much slower for $T \to 0$, $\om \to 0$
than in a Fermi liquid. At $T=0$, one obtains
$\Im\Sg(k_F,\om) \propto |\om|^{1-\eta}$, where $\eta$ is a
positive exponent which is relatively small for systems with
short-range interactions. 
The spectral function $A(k,\om)$ exhibits a single peak centered 
at $\om = 0$ for $k = k_F$, and a split peak at finite energy 
for $k$ away from the Fermi point.\cite{MS}
This peak splitting is not related to pseudogap behavior, 
it is rather due to spin-charge separation.
Applying our fRG scheme to a one-dimensional Luttinger liquid
system at weak coupling yields finite renormalized interactions, 
due to a well-known cancellation of singularities in the 
particle-particle and particle-hole channel. 
For the self-energy as computed from the flow equation (4) in
one dimension this implies a linear frequency dependence,
$\Im\Sg(k_F,\om) \propto |\om|$, yielding a single broadened
peak in $A(k_F,\om)$ as expected, but missing the anomalous
exponent $\eta$. The latter could be captured by feeding the
self-energy back on the right hand side of its flow equation,
most simply via a wave function renormalization, as described
for the one-particle irreducible fRG in Ref.\ \onlinecite{HS}.
The situation is different if umklapp scattering (so-called
$g_3$-terms) are relevant, as in the one-dimensional Hubbard
model at half-filling. In that case renormalized interactions
grow strong and a charge gap opens, leading to a gap in the
spectral function $A(k_F,\om)$, which is clearly signalled 
by a suppression of spectral weight at low frequencies within 
the fRG scheme.

Our results agree only partially with those obtained by Katanin
and Kampf,\cite{KK} who computed the full frequency dependence
of the self energy from the one-particle irreducible version of
the fRG,\cite{SH} with a truncation of the exact hierarchy of
flow equations at the same order as in our calculation.
For a Fermi momentum very close to a van Hove point they
obtained a single peak in $\Im\Sg$ resulting in a double peak
in the spectral function, in agreement with our calculation.
However, for Fermi momenta at some moderate distance
from the van Hove point they found a double peak in $\Im\Sg$
leading to a three-peak structure in the spectral function,
which we do not observe in our results. 
In both approaches, Wick-ordered as well as one-particle irreducible
fRG, the peaks in $\Im\Sg$ are generated by strongly enhanced 
effective interactions, and are thus most pronounced at temperatures
close to the instability scale $T^*$.
The different structure of $\Im\Sg$ obtained from the two fRG
versions can be traced back to a different division of momentum
integrals in high and low energy modes. In both cases the
energy variable $\om$ is related to the excitation energies on
internal lines: $\om = \xi_{\bp} + \xi_{\bp'} - \xi_{\bq}$.
In the Wick-ordered scheme all excitation energies are at or
below the cutoff, that is $|\xi| \leq \Lam$. As a consequence,
the flow for $\Im\Sg^{\Lam}(\bk,\om)$ with $|\om| > 0$ stops
completely for $\Lam < |\om|/3$.
In a situation where the vertex $\Gam^{\Lam}$ grows strongly at 
very small scales, $\Lam \to 0$, the imaginary part of the
self energy therefore receives the corresponding large contributions
only for small frequencies, which leads necessarily to a
single peak with maximal height at $\om = 0$.
By contrast, in the two-loop diagram generating the flow of
$\Im\Sg^{\Lam}$ in the one-particle irreducible scheme one
excitation energy is at scale $\Lam$, the second at a scale
$\Lam' > \Lam$, and the third one \emph{above} the scale $\Lam'$,
where $\Lam'$ is the scale of the vertex $\Gam^{\Lam'}$ in the
diagram. Note that the two-loop diagram for the self energy
in the one-particle irreducible fRG is non-local in the cutoff.
The crucial point is that not all excitation energies are
bounded by the cutoff.
Therefore, the flow of $\Im\Sg^{\Lam}(\bk,\om)$ does not
stop for $\Lam \ll |\om|$, and the strong enhancement of
$\Gam^{\Lam'}$ for $\Lam' \to 0$ does not necessarily enhance
$\Im\Sg^{\Lam}(\bk,\om)$ most strongly for the smallest
energies.
However, to obtain a double peak in $\Im\Sg$ with maxima at
finite frequencies, say $\om_{\pm}$, the contribution with at
least one excitation energy of order $\om_{\pm}$ must be
particularly large. This is possible only if the vertex is not
enhanced most strongly for cases with all ingoing and outgoing
momenta on the Fermi surface, but rather for some interaction
processes with at least one momentum away from the Fermi surface.
One should also keep in mind that
contributions from momenta away from the Fermi surface might be
overestimated in the one-particle irreducible fRG with a
momentum discretization by patches, since the dependence of the
vertex $\Gam^{\Lam}$ normal to the Fermi surface is weak only within 
a shell $|\xi_{\bk}| < \Lam$ in momentum space.

We finally note that the $Z$-factor defined as
$Z = [1 - \left.\partial_{\om}\Re\Sg(\bk_F,\om)\right|_{\om = 0}]^{-1}$
does not provide sufficient information to parametrize the low-frequency
dependence of the self energy, if the imaginary part develops a
low-energy peak.
Furthermore, $Z$ does not provide any information on the low-energy
weight in the spectral function in that case.
Since the slope of $\Re\Sg$ becomes positive near the hot spots,
the $Z$-factor does not even satisfy the inequality $0 < Z \leq 1$.

\section{Conclusion}

In summary, we have presented a weak-coupling calculation of the
self energy $\Sg(\bk,\om)$ for the two-dimensional Hubbard model
with a renormalization of effective interactions which fully
treats the interplay of particle-particle and particle-hole
channels.
Strong renormalizations of single-particle excitations via
self energy effects were observed for momenta near hot spots on
the Fermi surface, if the system is close to an instability
where the renormalized interactions are strongly enhanced.
In this case the quasi-particle peak in the spectral function
$A(\bk,\om)$ is replaced by a double-peak structure reminiscent
of a pseudogap. Further more, the fRG flow enables us to identify 
the correlations which are responsible for this feature, and thus offers
substantial insight into the underlying microscopic mechansim. 
Evidently, the perturbative truncation of the flow equations becomes 
more and more
uncontrolled in the regime of large renormalized interactions.
We can thus only conclude that single-particle excitations near
hot spots are strongly affected by interactions, but higher
order contributions to the self energy are expected to become 
important at a certain point.

The tendency to pseudogap formation found in the above weak-coupling 
calculation is restricted to a narrow temperature
scale and also to a narrow regime in momentum space, namely
close to the hot spots. In contrast, the pseudogap phenomena in 
hole-doped cuprate superconductors \cite{TS}
are observed in a much wider temperature range, and are seen almost
everywhere on the Fermi surface. This is undoubtedly due to the fact 
that the interaction in these materials is strong. Indeed, numerical 
results for the two-dimensional Hubbard model
at large $U$ obtained from quantum cluster approaches \cite{MJPH}
provide evidence for extended pseudogap behavior near
half-filling.

Our observation that for moderate interaction strength the pseudogap 
formation is favored at hot spots
is in accordance with recent results obtained by 
Kyung et al.\cite{KYU} on the basis of the two-particle 
self-consistent (TPSC) approach, as well as
with the findings by S\'en\'echal and Tremblay \cite{ST} from
cluster perturbation theory. In both references the focus is put on 
electron-doped cuprate superconductors, and it is argued
that these compounds may be more weakly coupled at optimal doping, 
as opposed to the hole-doped materials. Therefore, it seems more 
appropriate to compare our results with experimental data on 
electron-doped cuprates.
Note that at weak coupling the dominant correlations are strongly
affected by the the Fermi surface topology and its intersection 
with the umklapp surface, while no qualitative changes occur at
the transition from electron to hole-doped fillings. 
Hence, we may discuss qualitative properties of electron-doped
cuprates, although we computed for densities $n \leq 1$.
Unfortunately, the angular-resolved photemission (ARPES) data for
electron-doped cuprates 
available to this day are controversial. Very recently Claesson et 
al.~\cite{CLA} have presented measurements which do not show any signs  
of pseudogap behaviour, or other fingerprints of strong correlations 
in the $(\pi,\pi)$-transferring channel.  
On the other hand, earlier observations by Armitage et al.~\cite{ARM} 
are indeed partially consistent with our results. 
There, in a certain parameter range the Fermi surface vanishes 
around the hot spots where it would intersect the umklapp surface, 
and single-particle spectral weight is shifted 
to lower energies in that region. However, in these samples 
the qualitative picture changes as a function of doping. For a heavily 
underdoped system small electron pockets appear near $(\pi,0)$, 
while the remaining part of the would-be Fermi surface is destroyed, 
also along the zone diagonal. Toyama and Maekawa\cite{TOH} 
argue from calculations on a $t-t'-t''-J$ model 
that this may be due to the closeness to the 
antiferromagnetic phase, which is not as easily destroyed by doping 
as it is for hole-doped cuprates. Alternatively, 
this feature can be due to a small change in the repulsive 
interaction as a function of doping, as suggested in reference 
\onlinecite{KYU}. 
In both cases the interaction needs to be sufficiently large, 
and is outside the weakly coupled regime. 

A comprehensive understanding of cuprate superconductors and other
strongly correlated electron systems is still hampered by a lack
of theoretical methods and frequently also by difficulties in
obtaining reliable experimental data. 
At the present stage of its development the fRG can contribute
by offering a controlled and transparent treatment of electron 
correlations at weak coupling. This captures already a variety of
non-trivial phenomena, but misses genuine strong coupling physics,
such as the Mott transition. In the long run one may envisage
combining strong coupling methods for short range correlations with 
fRG schemes for the long range part.

\vskip 1cm

{\bf Acknowledgements:}
We are grateful to C. Honerkamp, A. Katanin, D. Manske, and 
M. Salmhofer for 
valuable discussions. This work has been supported by the DFG Grant
No.\ Me 1255/6-1.

\vskip 1cm


\vfill\eject


\begin{figure}
\center
\epsfig{file=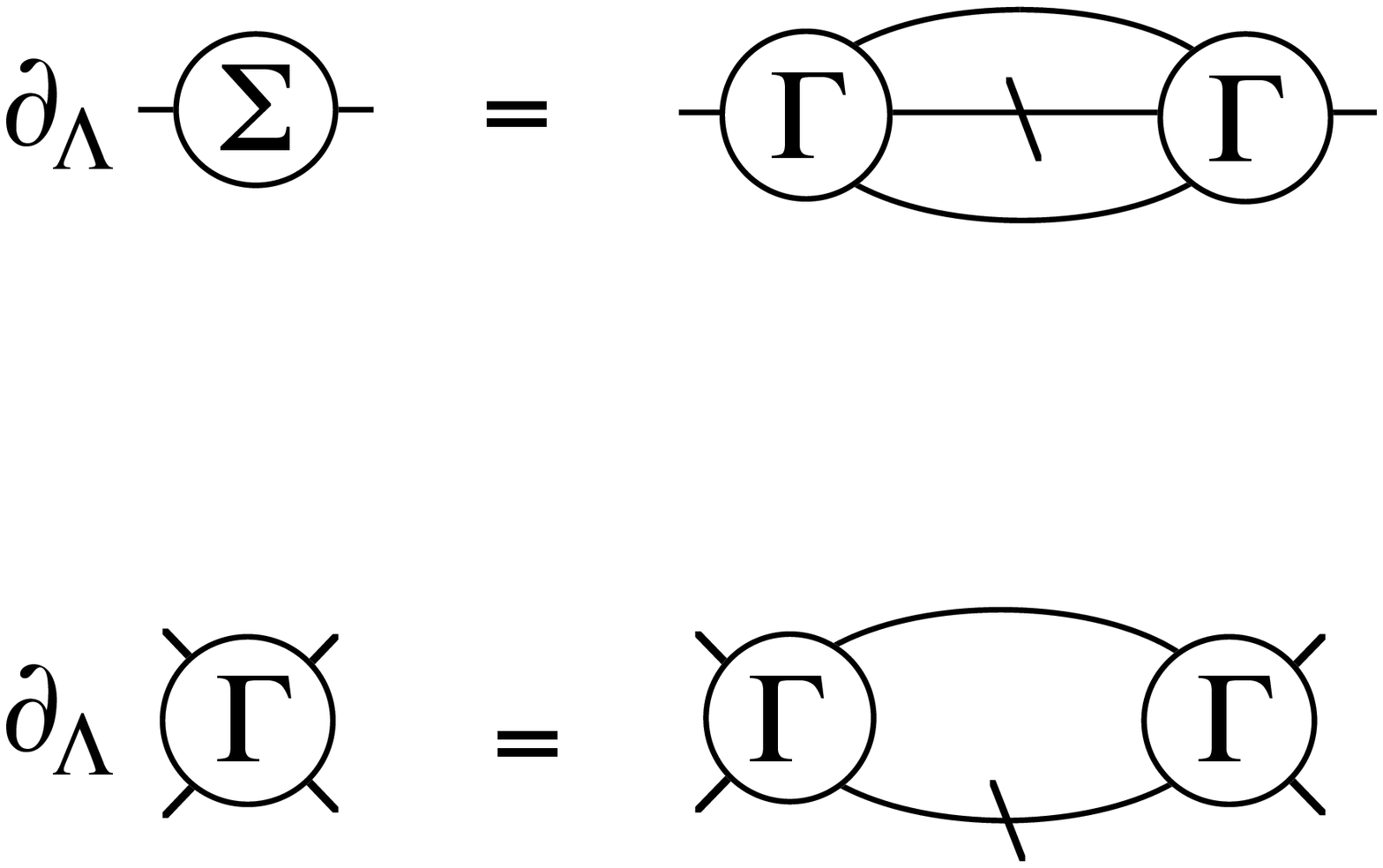,width=9cm}
\caption{Flow equations for the self energy $\Sg^{\Lam}$ and
 the two-particle vertex $\Gam^{\Lam}$, respectively;
 the internal lines without slash correspond to the bare propagator 
 $D^{\Lam}$, the lines with a slash to its $\Lam$-derivative 
 $\partial_{\Lam} D^{\Lam}$.}
\end{figure}

\begin{figure}
\center
\vskip 1cm
\epsfig{file=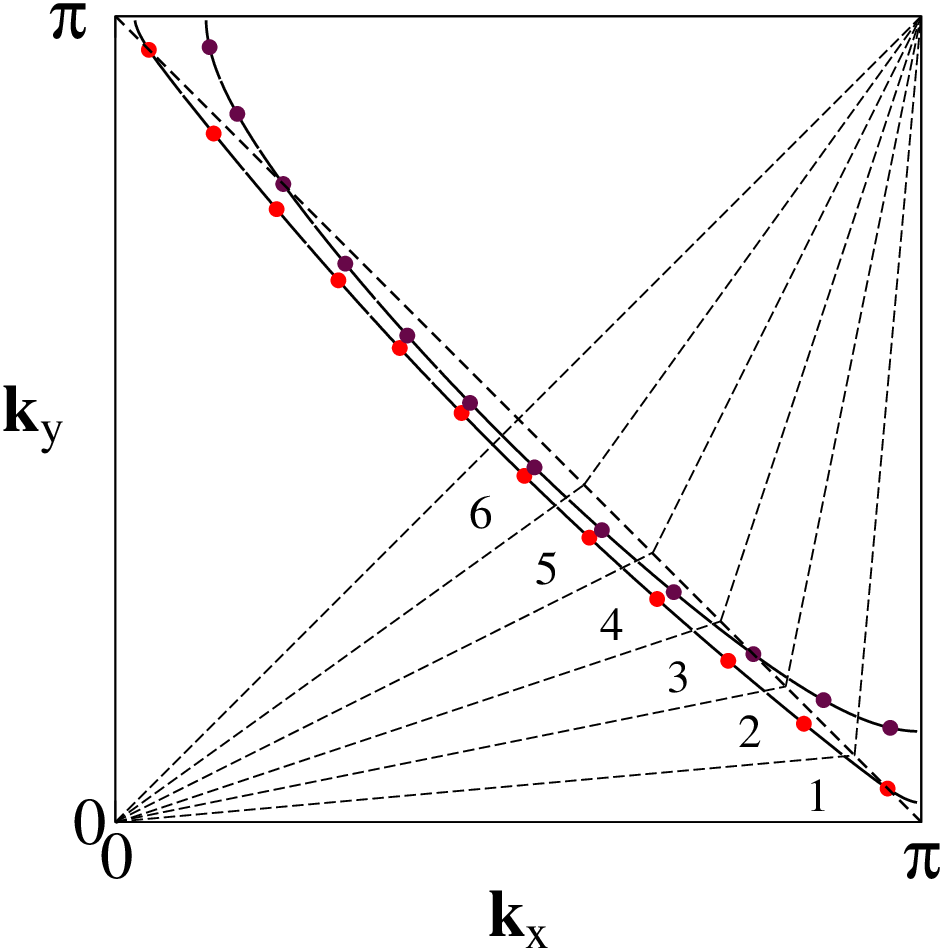,width=6cm}
\caption{Fermi surfaces (solid lines) for $n=0.92$ and $n=1$ 
 in the first quarter of the Brillouin zone, Fermi points used
 for the discretization of momentum dependences,
 and the first 6 momentum space patches
 (confined by broken lines from $(0,0)$ to $(\pi,\pi)$).
 The umklapp surface (broken line from $(0,\pi)$ to $(\pi,0)$)
 intersects the Fermi surface at the so-called hotspots.}
\end{figure}

\begin{figure}
\center
\epsfig{file=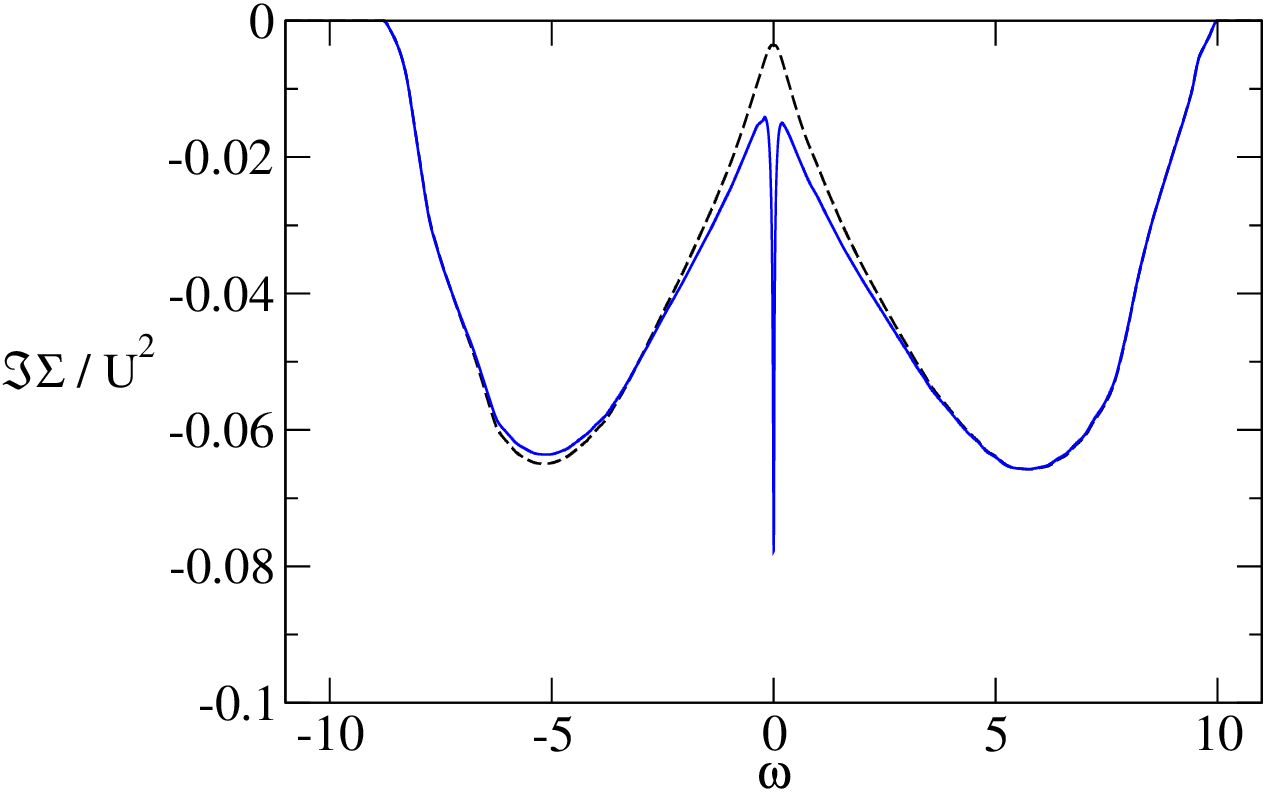,width=7cm}
\caption{Imaginary part of the self energy $\Im\Sg(\bk_F,\om)$
 normalized by $U^2$ as a function of $\om$ for $n=0.92$ and
 $\bk_F$ on the hot spot (patch 1, see Fig.\ 2).
 The result from the fRG (blue solid line) is compared to the
 result from second order perturbation theory (black dashed
 line).}
\end{figure}

\begin{figure}
\center
\vskip 1cm
\epsfig{file=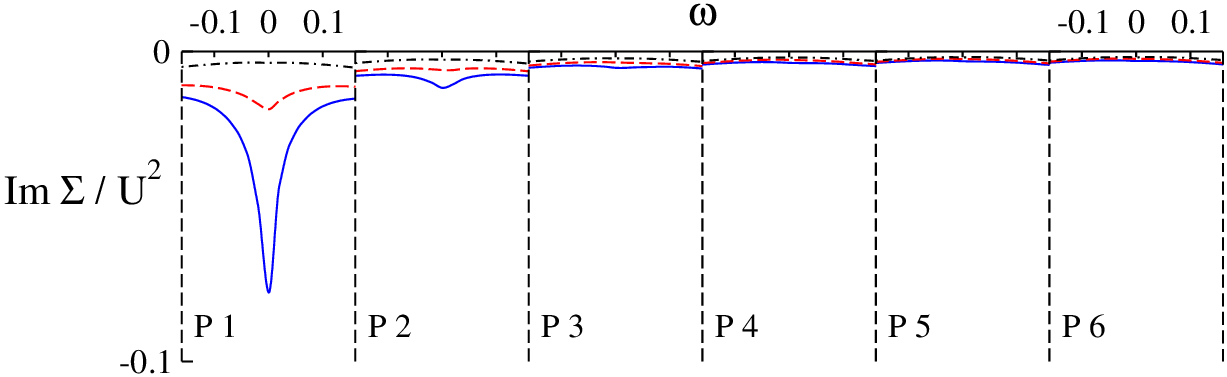,width=12cm}
\vskip 1cm
\epsfig{file=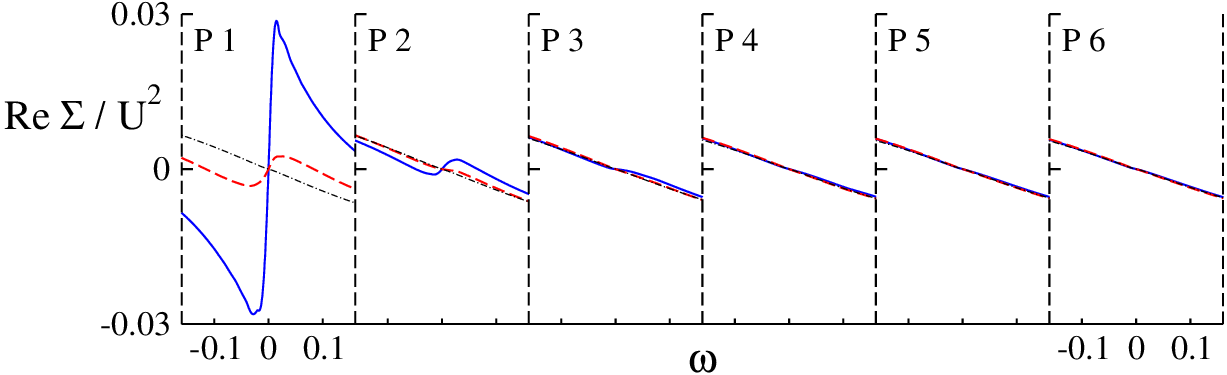,width=12cm}
\caption{$\Sg(\bk_F,\om)$ normalized by $U^2$ at low energies
 $\om$ for $n=0.92$ and $\bk_F$ in patches $1,2,\dots,6$.
 Two results from the fRG with $\Gam^{\Lam}_{\rm max} = 150$
 (blue solid lines) and $\Gam^{\Lam}_{\rm max} = 15$ (red dashed
 lines) are compared to each other; the former is obtained by
 choosing $U=1.815$, the latter for $U=1.685$.
 The results from second order perturbation theory for $U=1.815$
 (black dash-dotted lines) are also shown for comparison.}
\end{figure}

\begin{figure}
\center
\vskip 1cm
\epsfig{file=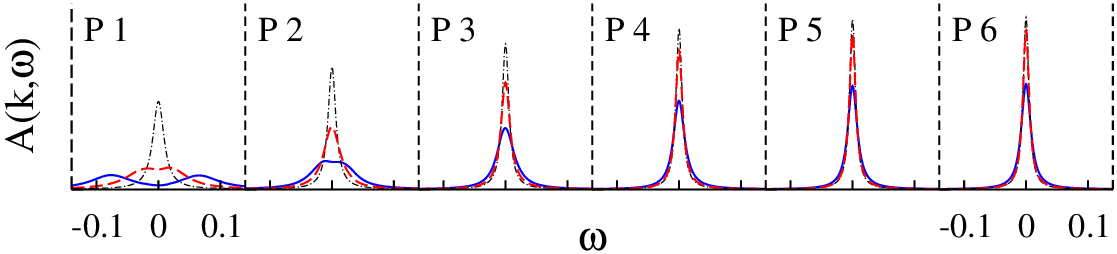,width=12cm}
\caption{Spectral function $A(\bk_F,\om)$ at small frequencies $\om$
 as obtained from the self energy results presented in Fig.\ 4.}
\end{figure}

\begin{figure}
\center
\vskip 1cm
\epsfig{file=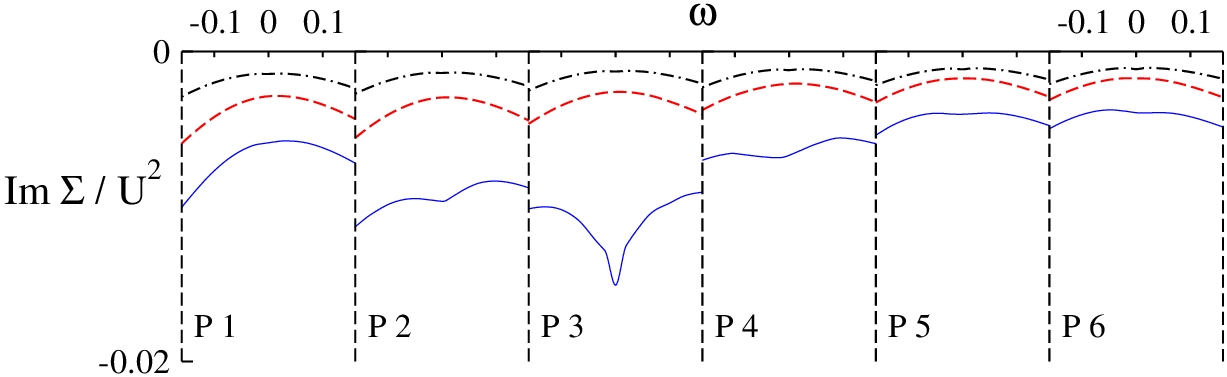,width=12cm}
\vskip 1cm
\epsfig{file=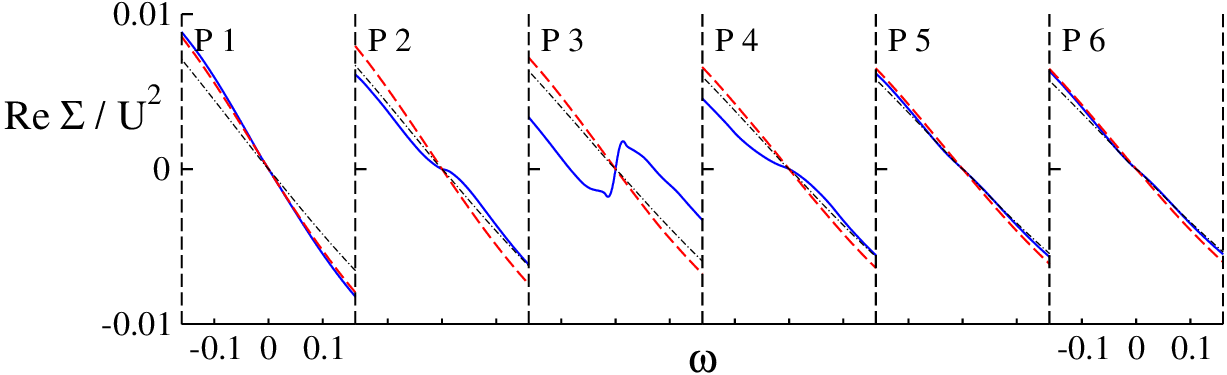,width=12cm}
\caption{$\Sg(\bk_F,\om)$ as in Fig.\ 4 but for $n=1$.}
\end{figure}

\begin{figure}
\center
\vskip 2cm
\epsfig{file=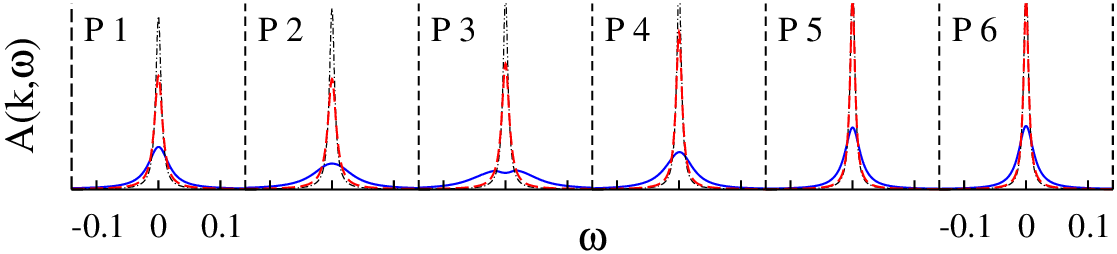,width=12cm}
\caption{Spectral function as in Fig.\ 5 but for $n=1$.}
\end{figure}

\end{document}